# INTERNATIONAL TOURISM AND GLOBAL BIODIVERSITY RISKS

Yingtong Chen[a], Fei Wu[b], Dayong Zhang[b], Qiang Ji[c]

[a] *School of Economics and Management, China University of Petroleum (East China), Qingdao, China*

[b] *Research Institute of Economics and Management, Southwestern University of Finance and Economics, Chengdu, China*

[c] *Institutes of Science and Development, Chinese Academy of Sciences, Beijing, China*

**Abstract:**

The impact of international tourism on biodiversity risks has received considerable attention, yet quantitative research in this field remains relatively limited. This study constructs a biodiversity risk index for 155 countries and regions spanning the years 2001 to 2019, analysing how international tourism influences biodiversity risks in destination countries. The results indicate that the growth of international tourism significantly elevates biodiversity risks, with these effects displaying both lagging and cumulative characteristics. Furthermore, spatial analysis shows that international tourism also intensifies biodiversity risks in neighbouring countries. The extent of its impact varies according to the tourism model and destination. In addition, government regulations and international financial assistance play a crucial role in mitigating the biodiversity risks associated with international tourism.

1. INTRODUCTION

For a long time, tourism has provided valuable development opportunities and economic benefits for underdeveloped regions. However, it also imposes "ecological costs", primarily in the form of biodiversity loss (van der Duim & Caalders, 2002). This is especially true for international tourism, as international tourists tend to engage in a wider range of activities. Due to lower environmental adaptability and awareness during their travels, the "ecological costs" of international tourism are generally higher than those of domestic tourism. Arguably, the negative impact of tourism on local ecosystems may outweigh its economic benefits.

According to the news by the United Nations World Tourism Organization (UNWTO), global tourism recovery in 2022 reached over 60% of pre-pandemic levels, with a rapid resurgence of international tourism in many countries and regions. This rapid growth has already started to have negative impacts in some ecologically sensitive areas (Primack, Bates, & Duarte, 2021). UNWTO predicts that the trend will continue in the coming years (UNWTO, 2024), placing a heavy burden on global ecosystems. Given that biodiversity is vital for maintaining the health and stability of ecosystems and a key tourism asset, examining the relationship between international tourism and biodiversity risks is crucial for ensuring the long-term sustainability of tourism.

The threat of international tourism to biodiversity manifests in various ways. The construction of tourism infrastructure has led directly to the clearance of native land vegetation and changes in the hydrology and soil of the areas adjacent to tourist destinations (Gogarty et al., 2018; Gül, 2023). Tourism-related vandalism, such as the overuse of waste, wetland clearance, sunscreen use, and commercial mining, has significantly increased the degradation of the international waters (Cowburn, Moritz, Birrell, Grimsditch, & Abdulla, 2018; Lin, Zeng, Asner, & Wilcove, 2023). Wilderness activities, such as camping, biking, and off-road driving, highly overlap with wildlife habitats, and the ensuing trampling on wild plants, picking flowers or fruits, spreading viruses, or "preventive" killing for safety threatens local biodiversity (Di Minin et al., 2021; Gogarty et al., 2018). Although tourism revenue, when appropriately used for local ecological conservation, has the potential to mitigate the negative impacts of tourism development (Ali, Yaseen, Anwar, Makhdum, & Khan, 2021), the profit-driven nature of the market often overlooks the ecological losses as a negative externality.

Despite its importance, existing research on the relationship between international tourism and biodiversity risks is largely limited to qualitative analysis, lacking detailed and robust quantitative evidence. Our study first adopts the Pressure-State-Response framework from existing literature to construct a comprehensive biodiversity risk index, enabling a scientific and systematic measurement of biodiversity risks across countries over time. Second, this study uses econometric analysis to examine the extent and characteristics of the impact of international tourism on biodiversity risks at tourism destinations. We aim to explore the ecological effects of international tourism development from both national and regional perspectives, providing further evidence for assessing these impacts in different natural and geographical contexts. In addition to this, we investigate potential risk mitigation strategies in three dimensions: international financial assistance, government environmental protection, and public attention, aiming to provide universal recommendations for addressing the ecological challenges associated with international tourism development.

The main contributions of this study are threefold: First, this study constructs a worldwide biodiversity risk index covering nearly 20 years, which overcomes the limitation of existing biodiversity risk indices for specific regions, and thus allowing us to conduct studies at a global scale. Second, this study explores the temporal and spatial characteristics of the impact of international tourism on biodiversity risks. It highlights the possible cross-country spillovers. By categorizing countries into regions, we offer new insights for regional cooperation in biodiversity protection in the context of tourism development. Third, although international organizations have proposed detailed action plans for achieving sustainable tourism, there is a lack of research on the effectiveness of these plans. This study empirically examines how to mitigate the impact of international tourism on biodiversity risks, offering evidence for preventing biodiversity risks in the development of international tourism.

## 2. LITERATURE REVIEW

Back in the 1970s, Clifford and Stephenson (1975) and Whittaker (1977) start to define methods for measuring species richness. It is the simplest indicator of biodiversity, used to estimate the number of different species in a specific area. However, the accuracy of this measurement is highly dependent on the sampling scale used by researchers in the studied area and is significantly limited by available manpower and resources. As a result, this method is difficult to apply at the national or global level.

In 1979, the widely used ecosystem health assessment model, the Pressure-State-Response framework, was introduced (Zheng & Huang, 2023). Following this framework, Stork and Samways (1995) create a composite index to assess the threat level to commercial tree species. Reyers, van Jaarsveld, McGeoch and James (1998) expand their work to create a national-level biodiversity risk index. The index, which considers both internal species populations and external environmental factors, has been widely recognized for providing a more comprehensive and accurate measure of biodiversity risk compared to simple changes in quantity or percentage (Molotoks, Kuhnert, Dawson, & Smith, 2017; Mozumder & Berrens, 2007). However, due to limited early data, this index lacks temporal continuity and could only reflect the previous state of biodiversity risk.

At the global level, two widely used indices are the Living Planet Index and the Red List Indices. Established in 1997, the Living Planet Index seeks to assess global biodiversity changes by tracking fluctuations in animal diversity and abundance (Loh et al., 2005). By continuously monitoring the population across various species worldwide, the Living Planet Index reports on biodiversity trends for different species and continents. However, it has not yet been refined to the national level. Additionally, this index primarily describes temporal changes in species instead of the actual biodiversity differences between species and regions. Butchart et al. (2004), when assessing global ecosystem degradation trends based on the Red List Indices, highlight that their updating process is rudimentary and subject to delays.

As an alternative to the direct use of habitat data, Giglio, Kuchler, Stroebel and Zeng (2023) develop the Biodiversity Risk News Index by analyzing articles from The New York Times. However, the data from The New York Times creates a clear geographical limitation, making it difficult to assess biodiversity risks in countries outside the United States. Given that biodiversity conservation is a critical global issue, the need for creating an international comparable measure at the national level is intensifying.

Turning to the issue of the impact of international tourism on biodiversity risks, previous literature has explored it with both qualitative and quantitative approaches. In qualitative research, some scholars have pointed out that even ecotourism is believed to have potential adverse effects on biodiversity (Gössling, 1999). These negative impacts may arise from waste disposal, infrastructure development, etc., with coastal and mountainous areas being particularly affected (Habibullah, Din, Chong, & Radam, 2016).

A typical approach to quantitatively studying this issue is by measuring the impact of international tourism on biodiversity through a "dose-effect" relationship (van der Duim & Caalders, 2002). Habibullah, Din, Chong and Radam (2016) explore the relationship between tourism and biodiversity and found that the increase in the number of international tourists exacerbated biodiversity loss. Chung, Dietz and Liu (2018) point out that the causal feedback between tourism and biodiversity may be highly complex. Yang, Xu, Pan, Chen and Zeng (2024) find that the expansion of human activities has imposed negative impacts on natural habitats globally. However, due to the complexity of defining biodiversity, the research conducted so far has not reached a consistent conclusion. Based on this strand of literature, we propose Hypothesis 1:

**H1: International tourism exacerbates biodiversity risks in destination countries.**

The most evident negative impacts of tourist activities on biodiversity include trampling of wild plants, fruit collection, intentional hunting and fishing, and the "preventive" killing of wildlife for safety reasons (Kuenzi & McNeely, 2008). Additionally, the unchecked development of infrastructure, excessive water use, and ongoing noise pollution further reduce the habitat space for local species (Ahmed et al., 2022). Although ecosystems possess a certain level of resilience to maintain equilibrium when disturbed, their ability to withstand disturbances and the speed at which they return to balance depend on the intensity and nature of the disturbance (Bao & Zhang, 2018). The greater the disturbance, the longer it takes for local biodiversity to recover, and in some cases, the ecosystem may never return to its original state. More critically, the loss of biodiversity sets off a cascade effect, leading to prolonged biodiversity risks. Increasing evidence suggests that the ongoing decline in biodiversity leads to the loss of critical ecological and socio-economic services, and the irreplaceability of ecological functions results in cascading changes within ecosystems (Säterberg, Sellman, & Ebenman, 2013; Dirzo et al., 2014). In general, biodiversity risks can accumulate over time, and thus, we propose Hypothesis 2:

**H2: The effects of international tourism on biodiversity risks endure over time and demonstrate cumulative impacts.**

The unrestricted nature of material cycles and energy flows within ecosystems allows ecological changes to transcend administrative boundaries (Yang et al., 2023). Human impacts on biodiversity are not confined to a single location but often trigger significant ripple effects in surrounding regions. For instance, Defeo et al. (2009) highlight that coastal conservation actions can effectively preserve biodiversity while generating beneficial spillover effects on neighboring beach habitats. Similarly, Brodie et al. (2023) find that ecological reserves positively influence biodiversity in adjacent areas. Wang, Guo and Li (2023) demonstrate that while economic aid increases environmental pressures within recipient countries, it also imposes burdens on neighboring nations due to regional linkages. Building on these findings, we hypothesize that international tourism not only directly threatens biodiversity in destination countries but also propagates ecological challenges to neighboring nations, thereby amplifying their biodiversity risks. Thus, we propose Hypothesis 3:

**H3: International tourism's impact on a country's biodiversity risks can spillover to neighboring nations.**

3. RESEARCH DESIGN

*3.1 Model specification*

To quantify the impact of international tourism on biodiversity risk, we construct a two-way fixed effects model, as shown in Equation (1).

$$lnBR_{it} = \alpha_1 + \alpha_2 lnTourist\_per_{it} + \Gamma Control_{it} + \eta_i + \nu_t + \varepsilon_{it} \qquad (1)$$

To avoid the potential impact of skewed data distribution on results, we take natural logarithm on all variables. Here, $lnBR_{it}$ represents the natural logarithm of Biodiversity Risk Index for country *i* in year *t*, $lnTourist\_per_{it}$ is the natural logarithm of international tourist arrivals per capita for country *i* in year *t*, $Control_{it}$ is the vector group of control variables that may affect biodiversity risk, $\eta_i$ and $\nu_t$ are country and year fixed effects, respectively, and $\varepsilon_{it}$ is the random error term.

To investigate the temporal characteristics of the impact of international tourism on biodiversity risk, we examine how the international tourism development influences biodiversity risk outcomes in future periods, as shown in Equation (2).

$$lnBR_{it} = \alpha_1 + \alpha_2 lnTourist\_per_{i,t-\tau} + \Gamma Control_{it} + \eta_i + \nu_t + \varepsilon_{it} \qquad (2)$$

Given the significant cross-regional mobility and spillover effects of international tourism, its impact on biodiversity often extends beyond the boundaries of a single country. Therefore, studying the effects of international tourism on biodiversity requires accounting for spatial spillover effects to comprehensively reveal its ecological impacts. The general form of a spatial econometric model is as follows:

$$\begin{cases} y = \rho W y + X\beta + \theta W X + \varepsilon \\ \quad\quad\varepsilon = \lambda W \varepsilon + v \end{cases} \quad (3)$$

Where $Wy$ represents the spatially lagged dependent variable, indicating how the dependent variable in one region is influenced by that in neighboring regions. Similarly, $WX$ denotes the spatially lagged explanatory variable, capturing the effect of explanatory variables in adjacent regions on the dependent variable in the focal region. The parameter $\rho$ is the spatial lag coefficient, measuring the spatial dependence of the dependent variable, while $\theta$ represents the spillover effect of the explanatory variables.

When $\lambda = 0$, the model is classified as a Spatial Durbin Model (SDM); when $\lambda = 0$ and $\theta = 0$, it simplifies to a Spatial Autoregressive Model (SAR); and when $\rho = 0$ and $\theta = 0$, it becomes a Spatial Error Model (SEM). Based on the likelihood ratio (LR) test (Supplementary Table 6), the Spatial Durbin Model is found to be the most appropriate model for this panel dataset. Consequently, we construct the Spatial Durbin Model according to Equation (3), incorporating specific variables as outlined in Equation (4).

$$lnBR_{it} = \xi + \rho \sum_{j=1}^{N} w_{ij} lnBR_{jt} + \gamma lnTourist\_per_{it} + \delta \sum_{j=1}^{N} w_{ij} lnTourist\_per_{jt} + \Gamma Control_{it} + \eta_i + v_t + \varepsilon_{it} \quad (4)$$

In Equation (4), $\sum_{j=1}^{N} w_{ij} lnBR_{jt}$ is the spatial lag of the dependent variable, $\sum_{j=1}^{N} w_{ij} lnTourist\_per_{jt}$ is the spatial lag of the explanatory variable, $\rho$ and $\delta$ are the corresponding regression coefficients, and $w_{ij}$ is the spatial weight matrix. We use the national adjacency spatial weight matrix and the national boundary linkage spatial weight matrix. Other variables are set the same as in Equation (1).

Furthermore, this paper builds an extended model by incorporating interaction terms to identify possible pathways for mitigating the impact of international tourism on biodiversity risk. The specific model is shown in Equation (5).

$$lnBR_{it} = \lambda + \beta C\_lnTourist\_per_{it} + \vartheta C\_Measure_{it} + \sigma C\_lnTourist\_per \times C\_Measure_{it} + \Gamma Control_{it} + \eta_i + v_t + \varepsilon_{it} \quad (5)$$

Here, $Measure_{it}$ represents possible mitigation measures, and other variables are set as in Equation (1). To ensure the comparability of the coefficients of the main variables in Equations (1) and (5), we follow the model specification of Balli and Sørensen (2013) by centring the explanatory variables or subtracting the mean from each variable (denoted as "C_") (Schielzeth, 2010).

*3.2 Variable selection*

*3.2.1 Measurement of biodiversity risks.* As concerns over biodiversity loss are growing, a range of methods for assessing biodiversity risks has been developed. (Butchart et al., 2004; Giglio, Kuchler, Stroebel, & Zeng, 2023; Stork & Samways, 1995). However, there remains a lack of relevant indicators to assess biodiversity risks across countries over continuous periods from a global perspective. Notably, as early as the 1980s, the Organisation for Economic Co-operation and Development (OECD) and the United Nations Environment Programme jointly introduced the foundational concept of an ecosystem assessment framework: the Pressure-State-Response framework. Due to its flexibility and comprehensiveness, this framework has been widely applied to assess the health of ecosystems such as wetlands, oceans, and rivers (Yang, Zhang, Sun, Liu, & Shao, 2021). In the Pressure-State-Response framework, *Pressure* refers to the stress or burden imposed on the ecological environment by human social and economic activities, *State* represents the condition of the ecosystem, and *Response* broadly encompasses the measures or actions taken to mitigate the adverse effects of human activities on the ecological environment. This framework is more comprehensive in evaluating the living environment of species compared to others (Supplementary Table 1).

The Biodiversity Risk Index constructed in this study builds upon the Pressure-State-Response framework, integrating the National Biodiversity Risk Assessment Index to systematically capture the interrelationships among Pressure, State, and Response. The National Biodiversity Risk Assessment Index, constructed by Reyers, van Jaarsveld, McGeoch and James (1998), has gained recognition in the fields of biodiversity evaluation and sustainable development (Molotoks, Kuhnert, Dawson, & Smith, 2017; Mozumder & Berrens, 2007). This indicator consists of three sub-indices: the *Pressure value*, *Stock value* and *Response value*, which directly correspond to the components of the Pressure-State-Response framework in terms of their definitions. Therefore, we adopt the construction method of the National Biodiversity Risk Assessment Index to develop the Biodiversity Risk Index used in this study, as shown in Equation (6). As the *Pressure value* (*PR*) increases or the *State* (*ST*) and *Response* (*RE*) values decrease, the Biodiversity Risk Index (*BR*) increases, indicating a higher level of biodiversity risk.

$$BR = \frac{PR}{RE \times ST} \qquad (6)$$

It is important to note that the National Biodiversity Risk Assessment Index is cross-sectional and can only be used to assess biodiversity risks for countries in 1997. due to data constraints, the proxy variables selected for constructing the indicator may lack precision. To address these limitations and to capture the changes in biodiversity risk over time, we adopt a series of indicators jointly developed by the authoritative biodiversity research organizations GEO BON and Map of Life: the Species Habitat

Index, Local Biodiversity Intactness Index, and Species Protection Index as proxy variables for the *Pressure value*, *State value*, and *Response value*, respectively.

These indicators are included in the official documents of the Convention on Biological Diversity as targeted metrics for implementing the First Draft of the Post-2020 Global Biodiversity Framework (Nicholson et al., 2021), ensuring the scientific rigor of the assessment. Based on the three indicators mentioned above, the Biodiversity Risk Index effectively covers all the dimensions of the National Biodiversity Risk Assessment Index while expanding on them (Supplementary Table 2). It addresses a broader range of aspects compared to existing biodiversity-related indices (Supplementary Table 3).

The Species Habitat Index reflects human impact on habitats—through factors like industrialization, deforestation, and pollution—by measuring changes in area, connectivity, and ecological integrity. We use its reciprocal in our analysis, so higher values indicate worse habitat conditions and greater pressure on species. The Local Biodiversity Intactness Index, which shows the remaining proportion of original biodiversity based on species abundance and disturbance levels; higher values mean better preservation. The Species Protection Index indicates local conservation efforts, with higher values reflecting stronger actions and lower species survival risks.

*3.2.2 Measurement of international tourism development.* To measure the development of international tourism, some existing studies use the number of inbound tourists as a proxy (Waqas-Awan, Rosselló-Nadal, & Santana-Gallego, 2021). However, the tourism arrival data may be affected by the size of population or scale of the destination country, thus we follow Anggraeni (2017) to standardize this variable and use arrivals per capita. That is to use the number of international tourist arrivals to the destination country divided by the population of the destination country. The data is collected from the World Bank Database[1] (e.g., Gozgor, Lau, Zeng, & Lin, 2019; Karabulut, Bilgin, Demir, & Doker, 2020) and covers the period from 2001 to 2019. In the subsequent empirical analyses, we use international tourist arrivals per capita as the main explanatory variable.

*3.2.3 Moderating Variables.* To ensure the long-term and sustainable development of tourism, it is crucial for organizations, governments, and local communities in destination countries to recognize the importance of sustainable tourism management. The tourism development model that exploits natural resources needs urgent changes. Therefore, this paper explores potential measures to mitigate the negative impacts of international tourism on biodiversity from three main perspectives: international financial assistance, government environmental protection, and public attention.

Biodiversity conservation projects are characterized by their strong public welfare nature, low returns, and long timelines (Booth, Arlidge, Squires, &

---

[1] Note that the original data for tourist arrival may include different definitions. Here we follow the World Bank and define it as "*the number of tourists who travel to a country other than that in which they usually reside, and outside their usual environment, for a period not exceeding 12 months and whose main purpose in visiting is other than an activity remunerated from within the country visited*" (the World Bank). A caveat to this is that when data is missing, the number of visitor (potentially cover different categories) is used.

Milner-Gulland, 2021). Additionally, many ecological protection actions inevitably restrict tourism activities, resulting in high opportunity costs (Zeng & Zhong, 2020). These factors contribute to a lack of initiative among governments and citizens in destination countries regarding biodiversity conservation.

External environmental aid can help balance the development of international tourism and ecological protection, encouraging less developed countries to prioritize biodiversity conservation. As climate change governance has become a priority in biodiversity management, this paper used climate change-related external development financing data from the OECD, adjusted for population size, as a proxy for the international financial support a country receives. While funding can increase local conservation efforts, governments play a crucial role in practical biodiversity protection (Wang, Feng, Liu, & Li, 2020). Thus, this paper examines the impact of government environmental efforts in mitigating tourism's negative effects on biodiversity, using the number of climate change policy clauses as a proxy for government action. Data is sourced from the 2021 Quality of Government Environmental Indicators dataset, published by the University of Gothenburg.

Additionally, public cooperation is vital for effective biodiversity conservation, even when local governments implement legal measures (Wang, Feng, Liu, & Li, 2020). This study assesses public concern for biodiversity by analyzing Google Trends data for the search term "Biodiversity" under the "Travel" category, using it as a proxy for public awareness in tourism destinations. A higher index suggests greater local awareness, potentially leading to increased participation in conservation efforts.

*3.2.4 Control variables.* The following factors that may affect a country's biodiversity risk are selected as control variables: the clean technologies and usage of renewable energy (Hishan et al., 2019), nitrous oxide emissions and agricultural greenhouse gas emissions (Boakes, Dalin, Etard, & Newbold, 2024), and industrial development (Watson et al., 2023). All the data are collected from the World Bank Database.

4. EMPIRICAL RESULTS

*4.1 Descriptive statistics*

To exclude the impact of COVID-19 on international tourism after 2020, this paper selects 2001-2019 as the sample interval. Due to the significant data gaps in some countries, and to ensure the validity of the results, we ultimately select 155 countries as research samples. **It is worth noting that the effective sample does not include island countries. Given that they do not share land borders with other nations, the construction of spatial weight matrix will be affected. These countries are typically small and thus excluding them does not affect the main results.** Descriptive statistics for all variables are shown in Table 1.

Table 1. Descriptive statistics.

| Variables | | N | Mean | S.D. | Min | Max |
|---|---|---|---|---|---|---|
| Dependent variable | *lnBR* | 2945 | 0.115 | 0.287 | 0.012 | 3.046 |
| Explanatory variable | *lnTourist_per* | 2945 | 5.152 | 1.834 | 0.419 | 10.176 |
| | *lnTourist* | 2945 | 14.390 | 2.032 | 7.973 | 19.199 |
| Moderators | *lnCli_pro* | 1923 | 9.930 | 2.791 | 0.145 | 16.159 |
| | *lnPolicy* | 2940 | 5.285 | 0.591 | 0.000 | 6.454 |
| | *lnMedia* | 2945 | 0.591 | 0.838 | 0.000 | 3.635 |
| Control variables | *lnACF* | 2945 | 0.213 | 0.128 | 0.029 | 1.030 |
| | *lnNO* | 2945 | 0.389 | 0.277 | 0.034 | 1.893 |
| | *lnRE* | 2945 | 3.061 | 1.217 | 0.010 | 4.599 |
| | *lnANOE* | 2945 | 7.862 | 2.062 | 1.832 | 12.790 |
| | *lnIVA* | 2945 | 3.251 | 0.388 | 1.715 | 4.362 |

Note: *BR* represents the Biodiversity Risk index we construct. *Tourist_per* is the country's inbound tourist arrivals per capita. *Tourist* is the country's inbound tourist arrivals. *Cli_pro* means the intensity of international financial assistance which is based on climate-related external development financing data. *Policy* represents the policy effectiveness, measured by the number of climate change policies. *Media* describes public attention, based on the search index of "Biodiversity" in the "Travel" section of Google Trends. *RE and ACF* represent the access to renewable energy (% of total final energy consumption) and clean cooking technologies (% of population) in countries. *NO* describes the nitrous oxide emissions (metric tons of $CO_2$ equivalent per capita). *ANOE* is *the* agricultural nitrous oxide emissions (thousand metric tons of $CO_2$ equivalent), indicating the intensity of human agricultural activities. *IVA* means the share of industry (including construction), value added (% of GDP), reflecting the intensity of human industrial activities. All variables are log-transformed, and multi-collinearity is tested among the key variables. The detailed results can be found in Supplementary Tables 4 and 5.

To clearly illustrate the distribution of biodiversity risk, we calculate the average Biodiversity Risk Index for each country over the sample period and presented the results in Figure 1. Fig 1(a) shows that countries with higher biodiversity risk are concentrated in Europe, Asia, and Africa. From Fig 1(b), it can be observed that regions with higher biodiversity risk are primarily located in the tropics, subtropics, and the northern temperate zone.

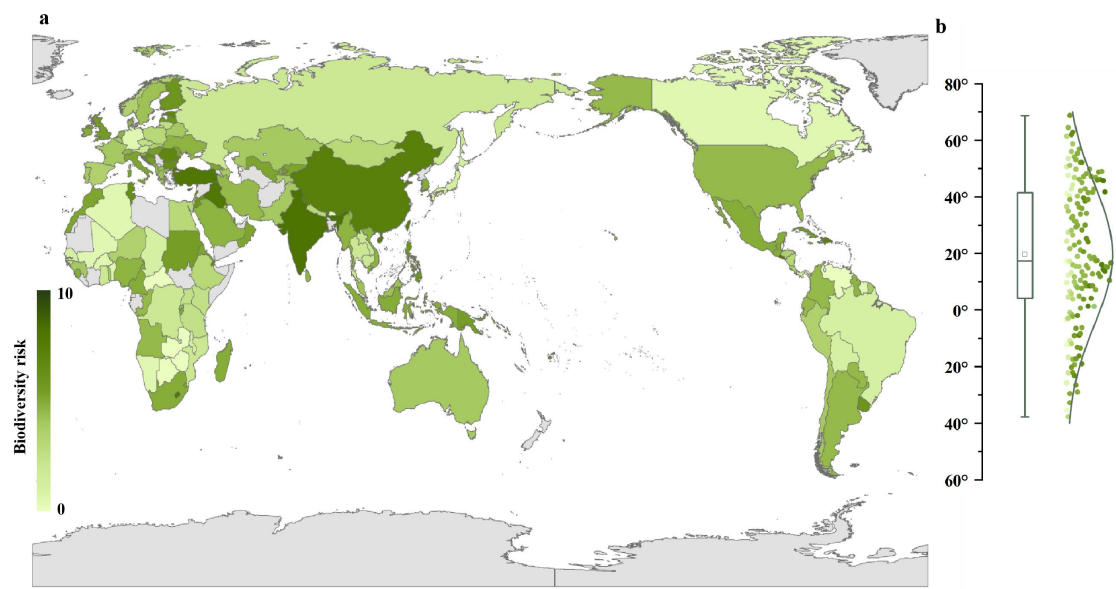

Fig 1. Global biodiversity risk distribution.

Note: Figure 1(a) displays the mean level of biodiversity risks across countries from 2001 to 2019. Figure 1(b) shows the distribution of biodiversity risk based on the latitude of each country's capital. Each point represents a country, with the colour coding consistent with the risk levels shown in Figure 1(a). Countries in grey: no data available.

### 4.2 The impacts of biodiversity risks

The baseline regression results are presented in Table 2. In Column (1)-(2), the core explanatory variable is the total number of inbound tourists. In Column (3)-(4), we use international tourist arrivals per capita[2]. Columns (1) and (3) show regressions without control variables, while columns (2) and (4) present the regression results with control variables included. The results show that the development of international tourism significantly increases the biodiversity risk in destination countries.

According to reports by the UNWTO, the global growth rate of international tourist arrivals is 4% in 2019, 6% in 2018, and 7% in 2017. In our study, an annual increase of 5%-10% in international tourist arrivals per capita leads to an increase in biodiversity risk by an average of 0.105%-0.210%. Although the numerical increase in biodiversity risk due to the development of international tourism appears relatively modest, it is important to contextualize this within the broader picture. Global species extinction rates are estimated to range between 0.01%-0.1% annually (Costello, May, & Stork, 2013). Given this extinction rate, the annual impact of international tourism on biodiversity risk cannot be overlooked.

Table 2. Impacts of international tourism on biodiversity risks

| $lnBR$ | (1) | (2) | (3) | (4) |
|---|---|---|---|---|
| $lnTourist$ | 0.014** | 0.015** | | |
| | (0.006) | (0.006) | | |
| $lnTourist\_per$ | | | 0.016* | 0.021** |
| | | | (0.009) | (0.009) |
| **$lnACF$** | | 0.237*** | | 0.245*** |
| | | (0.066) | | (0.066) |
| **$lnNO$** | | -0.296*** | | -0.295*** |
| | | (0.080) | | (0.080) |
| **$lnRE$** | | -0.046*** | | -0.046*** |
| | | (0.013) | | (0.013) |
| **$lnANOE$** | | 0.147*** | | 0.148*** |
| | | (0.030) | | (0.030) |
| **$lnIVA$** | | 0.024 | | 0.017 |
| | | (0.021) | | (0.021) |
| Constant | -0.084 | -1.133*** | 0.030 | -1.007*** |
| | (0.087) | (0.238) | (0.048) | (0.224) |
| Country/Year FE | YES | YES | YES | YES |
| N | 2945 | 2945 | 2945 | 2945 |

---

[2] Tourist arrivals per capita will be used in all following examinations as the main explanatory variable.

|       | Adj. $R^2$ | 0.665 | 0.674 | 0.665 | 0.675 |

Note: ***, **, and * indicate the 1%, 5%, and 10% significance levels, respectively. The robust standard errors are reported in parentheses.

*4.3 Temporal characteristics of the impact of international tourism.*

In this section, we examine the temporal trends in the impact of international tourism on biodiversity risks. We perform a regression with lagged international tourism data to assess its lasting impact on biodiversity risks. For the cumulative effect, we follow the approach outlined in the literature to calculate the sum of the coefficients representing the impact of international tourism development on biodiversity risk over future periods (Agarwal et al., 2021).

The results, presented in Table 3, indicate that the impact of international tourism on biodiversity risks remains relatively stable in the current year and the following year. However, the impact peaks in the second year, before gradually declining from the third to the fifth year and eventually becoming insignificant by the sixth year. Overall, the impact of international tourism development persists for up to five years but diminishes over time. This pattern aligns with the characteristic trends of systemic ecological disturbance and recovery processes (Dashti, Chen, Smith, Zhao, & Moore, 2024).

Table 3. Time trends in the impact of international tourism on biodiversity risks.[3]

| *lnBR* | (1) | (2) | (3) | (4) | (5) |
|---|---|---|---|---|---|
| L1_lnTourist_per | 0.019** (0.009) | | | | |
| L2_lnTourist_per | | 0.025*** (0.008) | | | |
| L3_lnTourist_per | | | 0.019*** (0.007) | | |
| L4_lnTourist_per | | | | 0.012** (0.006) | |
| L5_lnTourist_per | | | | | 0.008* (0.005) |
| Constant | -0.744*** (0.216) | -0.368* (0.194) | -0.230 (0.169) | -0.082 (0.144) | -0.067 (0.120) |
| Control variables | YES | YES | YES | YES | YES |
| Country/Year FE | YES | YES | YES | YES | YES |
| N | 2790 | 2635 | 2480 | 2325 | 2170 |
| Adj. $R^2$ | 0.702 | 0.754 | 0.811 | 0.868 | 0.914 |

Note: ***, **, and * indicate the 1%, 5%, and 10% significance levels, respectively. The robust standard errors are reported in parentheses.

Furthermore, we categorize countries by continent and analyse the characteristics of how international tourism impacts biodiversity risks over time, as illustrated in Fig 2. According to Fig. 2(a), there are significant differences in cumulative effects among countries in Asia, Europe, and North America, while the differences are relatively smaller among countries in Africa, Oceania, and South America. From the density plot, it is evident that the cumulative effects are mostly above zero, indicating that international tourism is likely to exacerbate the biodiversity risks in destination countries for years to come. Fig. 2(b) demonstrates that the development of international tourism has consistently heightened biodiversity risks across all continents. Based on the scatter plot distribution, the impact coefficients are predominantly positive, with their values showing a distinct convergence during the 3rd to 5th years in the future.

---
[3] Each lagged explanatory variable is modelled separately instead of pooling them together due to high level of autocorrelations, which leads to the issue of multi-collinearity. We do appreciate the comments and suggestions from the reviewer on this matter.

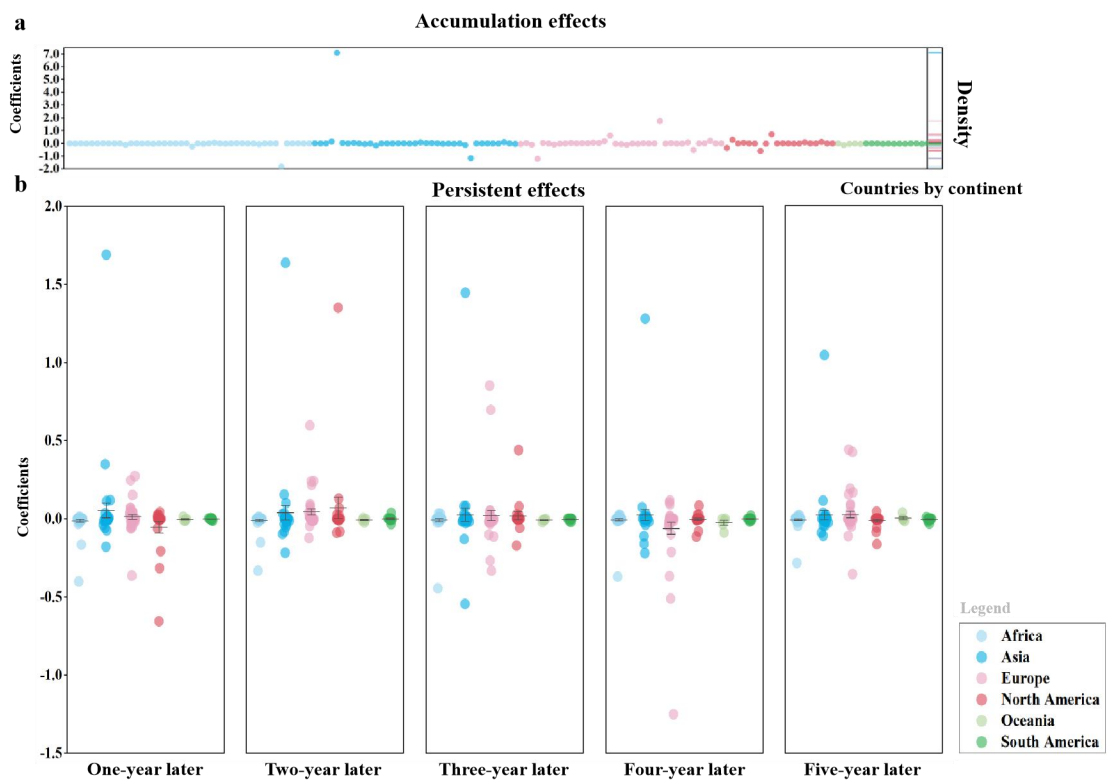

Fig 2. The cumulative effects and persistent effects of international tourism on biodiversity risks.

Note: The cumulative effect in Fig. 2(a) is calculated as the sum of the international tourism impact coefficients on biodiversity risks for each country over the next 1-5 years. A density plot on the right illustrates the distribution of the cumulative effect coefficients. Fig. 2(b) is organized into five columns, each representing the persistent effect of international tourism on biodiversity risks for the next one, two, three, four, and five years, respectively. Colours represent countries from different continents. Each point on the plot corresponds to the effect of international tourism on local biodiversity for a specific country within that continent. Error bars are used to represent the variability of these impacts across different continents.

### 4.4 Spatial characteristics of the impact of international tourism.

Due to the continuity of geographical areas and the interconnectedness of ecosystems (Hoang et al., 2023), biodiversity loss in one country can influence neighbouring countries through energy exchange between ecosystems. To explore the spatial characteristics of the global impact of international tourism on biodiversity risks, the Spatial Durbin Model is used (as shown in Table 4). Initially, we utilize a national spatial adjacency weight matrix, where the matrix element is assigned a value of 1 if two countries share a border and 0 otherwise. However, because national boundaries can be defined by various natural features such as rivers, lakes, and mountains, the spatial adjacency weight matrix may inadequately capture the flow of energy and cycling of materials between neighbouring ecosystems. To address this limitation, we also use data from the Correlates of War project to construct a national boundary linkage spatial weight matrix, which incorporates the types of national boundaries. This approach provides a more accurate representation of ecological interconnectedness between countries.

Using the national spatial adjacency weight matrix (shown in Column 1-3, Table 4), we find that an increase in international tourist arrivals per capita to one country significantly exacerbates the biodiversity risk in neighbouring countries. Specifically, a 1% increase in international tourist inflows per capita directly raises local biodiversity risk by an average of 0.023%, and results in an average increase of 0.024% in the biodiversity risk of neighbouring countries. Overall, a 1% increase in international tourism raises regional biodiversity risk by 0.047%. This coefficient is significantly higher than the one obtained in the basic regression, indicating that the negative impact of international tourism on biodiversity is amplified through regional interactions.

Moreover, using the national boundary linkage spatial weight matrix leads to another important finding: the higher the accessibility between two countries, the greater the spatial spillover effect of international tourism development on biodiversity risks in neighbouring countries. The coefficients of other control variables show little change compared to those in Columns 1-3, highlighting the robustness of the empirical results.

Table 4. Spatial spillover effects of international tourism on biodiversity risks.

| lnBR | National spatial adjacency weight matrix | | | National boundary linkage spatial weight matrix | | |
| --- | --- | --- | --- | --- | --- | --- |
| | Direct effect | Indirect effect | Total effect | Direct effect | Indirect effect | Total effect |
| lnTourist_per | 0.023** | 0.024* | 0.047*** | 0.024** | 0.020* | 0.044*** |
| | (0.009) | (0.012) | (0.016) | (0.009) | (0.011) | (0.015) |
| Control variables | YES | YES | YES | YES | YES | YES |
| Country/Year FE | YES | YES | YES | YES | YES | YES |
| Sigma$^2$ | | 0.025*** | | | 0.025*** | |
| | | (0.001) | | | (0.001) | |
| N | | 2945 | | | 2945 | |
| $R^2$ | | 0.009 | | | 0.009 | |

Note: ***, **, and * indicate the 1%, 5%, and 10% significance levels, respectively. The robust standard errors are reported in parentheses.

Research has indicated that the transmission of biodiversity risks varies depending on geographical or climatic conditions (Lu et al., 2020). Therefore, we classify the sample countries based on their geographical location and temperature zone distribution to examine whether the development of international tourism in certain regions or ecosystem is more likely to cause biodiversity risk spillovers. The results are presented in Figure 3.

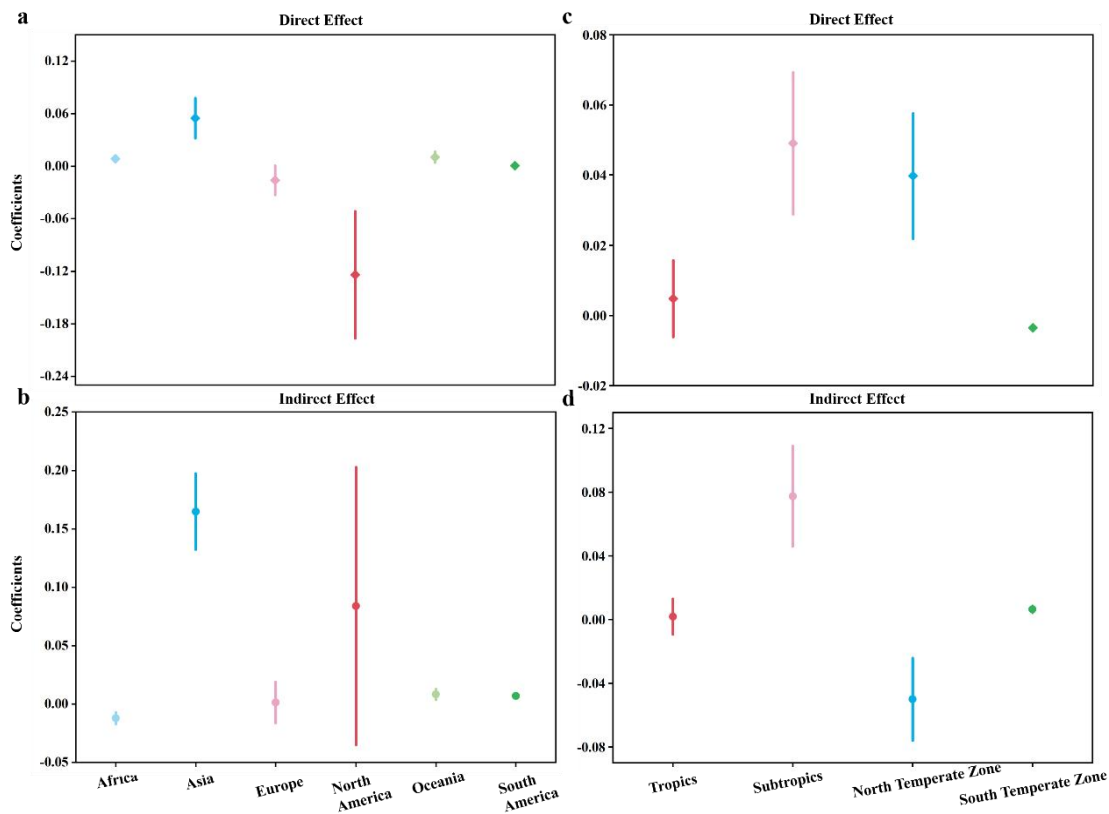

Fig 3. Differences in the spatial spillover effects of international tourism on biodiversity risks across regions

Note: Solid dots represent the coefficient values of the impacts, while vertical lines indicate the 95% confidence intervals of the coefficients. The spatial direct and indirect effects of international tourism on biodiversity risks are shown for countries classified by continent in Fig. 3(a) and 3(b), respectively. Fig. 3(c) and 3(d) display the spatial direct and indirect effects of international tourism on biodiversity risks for countries classified by thermal zone. For detailed regression coefficient results, please refer to Supplementary Tables 7 and 8.

An interesting conclusion can be drawn from Fig. 3(a): the development of international tourism has a significant mitigating effect on local biodiversity in North American countries. This may be because countries in North America, particularly the United States and Canada, have well-established systems of nature reserves and national parks (Buckley, 2002). Many of these reserves use tourism revenue to support biodiversity conservation and research, thereby creating a positive feedback loop in which international tourism contributes to ecological protection (Wight, 2001). As a result, international tourism in North America helps mitigate local biodiversity risks. Fig. 3(b) shows that the impact of international tourism on biodiversity risks in neighboring regions is most pronounced in Asian countries. This is largely due to Asia's extensive transboundary ecosystems with strong internal connectivity (Barletta et al., 2010; Neupane, Baral, Risch, & Campos-Arceiz, 2022), such as the Mekong River Basin. Biodiversity risks arising from international tourism in one country can easily spill over and affect neighboring countries.

The results of the spatial effect analysis after classifying countries by temperature zone, are shown in Figure 3(c) and 3(d). A comparison of the two figures shows that the impact of international tourism on both local and neighboring biodiversity risks remains relatively consistent in magnitude and direction for countries in tropical, subtropical, and southern temperate zones. However, the northern temperate zone demonstrates a distinct pattern. One possible explanation is that the northern temperate zone, characterized by its pleasant climate and distinct seasonal ecological landscapes, attracts tourists year-round, placing direct pressure on local ecosystems. However, countries in this region are also highly engaged in cross-border conservation initiatives and the export of environmental protection technologies (Khurshid, Huang, Cifuentes-Faura, & Khan, 2024; Thogmartin et al., 2022). These collaborative measures provide significant benefits to neighboring countries within the region, contributing to the mitigation of overall biodiversity risks.

*4.5 Heterogeneity of the impacts of international tourism on biodiversity risks.*

The biodiversity risks associated with international tourism development vary under different conditions. To investigate this, we analyse the heterogeneity in the impact of international tourism on biodiversity risks using the UNWTO classification of international tourists. The results are presented in Table 5.

As shown in Columns 1-2, business travelers significantly mitigate local biodiversity risks, whereas an increase in tourists visiting for personal reasons (mainly leisure and recreation) substantially heightens these risks. A possible explanation is that business travelers typically have shorter stays and tend to limit their activities to urban areas. Moreover, many business trips involve corporate engagements, which are often subject to environmental regulations and sustainability commitments (Rapaccini et al., 2020). In contrast, personal travel primarily revolves around relaxation (Pearce & Lee, 2005), with tourists being more likely to engage with natural environments, thereby posing a greater threat to local biodiversity.

The longer international tourists stay in destination countries, the more attractions they may visit, and the stronger impact they may have on local biodiversity risks (Pickering & Hill, 2007). We test this hypothesis by dividing international visitors into same-day trips and overnight stays[4] (Columns 3-4). Overnight stays significantly increase local biodiversity risks compared to same-day visits. This is because longer stays often involve visits to more tourist sites, leading to greater ecological damage (Belsoy, Korir, & Yego, 2012).

Moreover, the impact of international tourism on local biodiversity risks varies depending on the development level of the destination country. (Karani & Failler, 2020). We categorise the sample countries as developed or developing countries (Column 5-6), according to the International Monetary Fund. The results align with expectations, as developed countries are better equipped to achieve a positive balance between tourism and ecological needs (Waldron et al., 2013). In contrast, many developing countries, driven by economic growth objectives, often lack the capacity to effectively manage the ecological impacts of high tourist volumes (Goffi, Cucculelli, & Masiero, 2019), leading heightened biodiversity risks.

---

[4] The information on overnight/same-day visitors are taken from UNWTO. Please refer https://www.unwto.org/tourism-statistics/key-tourism-statistics for more information.

Table 5. Heterogeneity analysis.

| lnBR | Business | Personal | Same day | Overnight | Developed country | Developing country |
|---|---|---|---|---|---|---|
| | (1) | (2) | (3) | (4) | (5) | (6) |
| lnTourist_per | -0.023*** | 0.026** | 0.008 | 0.030*** | -0.041*** | 0.035*** |
| | (0.010) | (0.012) | (0.010) | (0.011) | (0.012) | (0.011) |
| Constant | -0.714*** | -0.912*** | -1.580*** | -1.155*** | -1.398*** | -0.809*** |
| | (0.281) | (0.288) | (0.412) | (0.246) | (0.506) | (0.253) |
| Control variables | YES | YES | YES | YES | YES | YES |
| Country/Year FE | YES | YES | YES | YES | YES | YES |
| N | 2168 | 2184 | 1448 | 2570 | 608 | 2337 |
| Adj. $R^2$ | 0.713 | 0.714 | 0.436 | 0.698 | 0.473 | 0.683 |

Note: ***, **, and * indicate the 1%, 5%, and 10% significance levels, respectively. The robust standard errors are reported in parentheses.

*4.6 Measures to mitigate the negative effects on biodiversity.*

To mitigate the impact of international tourism development on local biodiversity risks, we explore feasible solutions at three levels: international financial assistance, government environmental protection, and public attention (Aznarez et al., 2023; Echeverri et al., 2023; Seidl et al., 2021).

The results reveal that international financial assistance and government environmental protection significantly reduce the impact of international tourism on biodiversity risks. However, public attention to biodiversity does not have a significant mitigating effect. This outcome can be attributed to the following factors: First, public attention to biodiversity in the tourism sector often fails to translate into concrete conservation actions. Second, increased public awareness does not necessarily lead to improvements in infrastructure or management practices. In conclusion, raising public awareness of biodiversity risks alone is insufficient to meet the complex demands of biodiversity conservation. Substantial financial support and enhanced management capabilities are essential for effectively alleviating the biodiversity risks associated with international tourism.

Table 6. Possible ways to mitigate the risks of international tourism affecting

| lnBR | (1) | (2) | (3) |
|---|---|---|---|
| lnTourist_per | 0.017** | 0.002 | 0.020** |
| | (0.008) | (0.010) | (0.009) |
| C_lnTourist_per×C_lnCli_pro | -0.003*** | | |
| | (0.001) | | |
| C_lnCli_pro | -0.004* |  |  |
|  | (0.002) |  |  |
| C_lnTourist_per×C_lnPolicy |  | -0.046*** |  |
|  |  | (0.006) |  |
| C_lnPolicy |  | -0.244*** |  |
|  |  | (0.025) |  |
| C_lnTourist_per×C_lnMedia |  |  | -0.001 |
|  |  |  | (0.002) |
| C_lnMedia |  |  | 0.002 |
|  |  |  | (0.006) |
| Constant | -0.157 | -0.518** | -0.907*** |
|  | (0.251) | (0.220) | (0.217) |
| Control Variables | YES | YES | YES |
| Country/Year FE | YES | YES | YES |
| N | 1923 | 2940 | 2945 |
| Adj. $R^2$ | 0.846 | 0.688 | 0.674 |

Note: ***, **, and * indicate the 1%, 5%, and 10% significance levels, respectively. The robust standard errors are reported in parentheses.

## 5. ROBUSTNESS CHECK

To ensure that the results are not biased due to endogeneity issues, we employ several methods in this section to validate the robustness of main conclusions. First, we apply the Oster test to examine the impact of unobserved omitted variables on the results. As shown in Column (1) of Table 7, the delta value reported by the Oster test is less than 0, indicating that the regression results are robust (Oster, 2019).

Second, to address the potential endogeneity issue caused by reverse causality, we construct a Difference-in-Difference (DID) analysis to examine the impact of international tourism on biodiversity risks. Major sporting events, such as the Olympics, can attract international tourists to the host country, generating a long-term positive effect on the country's international tourism (Burgan & Mules, 1992; Solberg & Preuss 2007). Countries that hosted the Summer and Winter Olympics during the sample period can be used as a treatment group in the DID model. The results in Column (2) of Table 7 indicate that hosting a major sporting event leads to a significant increase in local biodiversity risks. Column (3) presents the DID results after Propensity Score Matching (PSM-DID). These findings further confirm that the robustness impacts of international tourism on biodiversity risks.

In addition, this study uses an instrumental variable approach and the two-stage least squares method (IV-2SLS) for addressing endogeneity problem. The instrumental variable is the number of world's intangible cultural heritage sites[5], regularly selected by the United Nations Educational, Scientific and Cultural Organization. Columns (4) and (5) of Table 7 present the first- and second- stage results using the instrumental variable (*lnHeritage*). As shown in Column (4), the instrumental variable is significantly and positively correlated with the number of international arrivals per capita. The fitted values of core explanatory variable (***H(lnTourist_per)***) continue to show a significant positive effect on biodiversity risks, demonstrating the robustness of the results. Moreover, the IV has passed both the identification and weak instrument tests, confirming its validity.

---

[5] Intangible cultural heritage attracts tourists, but it has no direct relationship with biodiversity risks, making it a good candidate of being a valid instrument.

Table 7. Robustness test results

| *lnBR* | OLS | DID | PSM-DID | IV-2SLS First-stage | IV-2SLS Second-stage |
|---|---|---|---|---|---|
|  | (1) | (2) | (3) | (4) | (5) |
| *lnTourist_per* | 0.021** |  |  |  |  |
|  | (0.009) |  |  |  |  |
| Game |  | 0.066** | 0.061** |  |  |
|  |  | (0.031) | (0.030) |  |  |
| *lnHeritage* |  |  |  | 0.074** |  |
|  |  |  |  | (0.037) |  |
| ***H(lnTourist_per)*** |  |  |  |  | **0.741*** |
|  |  |  |  |  | **(0.433)** |
| Constant | -1.007*** | -0.898*** | -1.039*** | 6.472*** | -5.592** |
|  | (0.224) | (0.217) | (0.244) | (0.500) | (2.825) |
| Control variables | YES | YES | YES | YES | YES |
| Country/Year FE | YES | YES | YES | YES | YES |
| N | 2945 | 2945 | 2604 | 2945 | 2945 |
| Adj. $R^2$ | 0.675 | 0.674 | 0.705 | 0.967 | 0.043 |
| Oster test (delta) | -0.510 |  |  |  |  |
| Kleibergen-Paap rk LM statistic |  |  |  |  | 4.08** |
| Stock-Wright LM S statistic |  |  |  |  | 43.53*** |

Note: ***, **, and * indicate the 1%, 5%, and 10% significance levels, respectively. ***H(lnTourist_per)* refers to the predicted value of *lnTourist_per* form the first stage regression.** The robust standard errors are reported in parentheses.

## 6. DISCUSSION

Tourism, as a socio-economic activity centered on human endeavors, has long been debated for its relationship with the ecological environment. The impacts of tourism on the environment span multiple research disciplines and various spatial scales (Liburd, Menke, & Tomej, 2024; Tajer & Demir, 2022). Despite extensive discussion, our understanding of the interaction between tourism and the ecological environment remains incomplete. Current studies on the tourism's ecological impacts tend to focus more on the "environment" than on the "biological" aspects (Baloch et al., 2023; Lee & Chen, 2021). However, biodiversity, as the foundation of ecological processes and ecosystem services, represents a critical yet underexplored area of research. Furthermore, international tourism, a significant driver of global population mobility, has unique cross-border effects on biodiversity (Hehir, Scarles, Wyles, & Kantenbacher, 2023). Its scope, intensity, and characteristics differ substantially from those of domestic tourism, but these distinctions have received limited attention.

Our study follows the Pressure-State-Response framework, a conceptual model grounded in system theory, to capture the interrelationships among various factors influencing biodiversity. Through this framework, we assess biodiversity risks at the national level and analyze the global impact of international tourism on biodiversity.

Our findings indicate that the expansion of international tourism significantly heightens biodiversity risks in destination countries, confirming what has been noted in previous studies (Lee & Chen, 2021). Dynamic analysis reveals that the impact of international tourism on biodiversity risks exhibits a lag effect, with its influence expected to intensify in the coming years. This phenomenon can be attributed to two key factors: the natural delay in biodiversity's response to environmental pressures (Watts et al., 2020), and the persistent impacts of tourism development activities (Tolvanen & Kangas, 2016). Additionally, the mismatch between the pace of tourism development and the timely implementation of biodiversity conservation measures (Hall, 2011) exacerbates the cumulative impact of international tourism on biodiversity risks in destination countries.

From a spatial perspective, our study shows that the effects of international tourism on biodiversity risks in destination countries extend to neighboring regions. This occurs because ecosystems do not recognize national borders (Westney, Piekkari, Koskinen, & Tietze, 2022). Tourism pressures in one area can threaten the ecological functions of surrounding regions. Conversely, effective ecological governance in one country can benefit neighboring regions. These findings emphasize the importance of strengthening international cooperation and aligning tourism development policies across countries.

## 7. CONCLUSIONS

While the international tourism brings economic benefits to destination countries, it also poses great risks to biodiversity locally and even on a larger scale. Previous literature lacks comprehensive, globally based quantitative studies on the relationship between international tourism and biodiversity risks, leaving temporal and spatial patterns underexplored. This study quantifies the relationship between international tourism and biodiversity risks, filling gaps in the theoretical framework on the ecological impacts of tourism.

First, a national biodiversity risk index is developed to quantitatively analyse the impact, temporal and spatial distribution, and heterogeneity of biodiversity risks driven by international tourism across 155 countries from 2001 to 2019. Empirical results can be summarized in four key points: (1) international tourism amplifies biodiversity risks in destination countries, with these risks showing both lagged and cumulative effects; (2) the impacts on biodiversity risks exhibit spillover effects that vary across regions; (3) the impacts differ significantly across the types and modes of tourism; (4) international financial assistance and government environmental protection can effectively mitigate the negative impacts of international tourism.

In a sense, the conflict between the development of international tourism and biodiversity is a "tragedy of the commons", which can only be resolved by establishing a "mutual coercion, mutually agreed upon" strategy. The results in this study are expected to contribute to the design of sustainable tourism policies. Specifically, when international organizations allocate financial aids, they should take the country's industrial structure into account, especially in destination countries with unique ecosystems or heavily dependent on natural resources. The financial aids should be linked to specific environmental goals and accompanied by monitoring mechanisms. Organizations should regularly assess the actual impact of the aids on biodiversity conservation and adjust their funding allocation strategies based on the feedback received.

Governments should design ecological monitoring programs tailored to regional conditions, refine land-use planning, and evaluate the environmental impacts of various tourism facilities after implementation. Meanwhile, they should develop mandatory tourism management plans for natural and mixed tourist areas and strengthen cooperation with other countries and organizations to share ecological monitoring data. Additionally, governments could establish comprehensive financing mechanisms to allocate tourism revenue towards biodiversity conservation.

While this paper provides valuable results for a critically important issue, the empirical study is limited in several aspects and deserves further investigation. First, more efforts should be given to find more accurate measures of biodiversity risk. Due to data availability, certain factors (e.g., genetic biodiversity) cannot be easily captured in the index, especially at the global level. Second, although we do not find significant role of public attention, it does not undermine the needs for more general participation to product the ecosystem. In addition, the current measure of using Google Trends reflects only a broader proxy, and thus unable to cover particular public issues (e.g., environmental activism). As an alternative, indicators based on surveys (e.g., Bernardo, Loupa-Ramos, & Carvalheiro, 2021) can be useful though it may only apply to a small set of countries. Furthermore, building on the findings of the temporal and spatial effects of tourism development on biodiversity risk, future research could explore a comprehensive spatial-temporal model that captures both effects, paying special attention to the possible interactive relationship over time and across regions.